\documentclass{elsart}

\usepackage{natbib}

\usepackage{graphics}
\usepackage{graphicx}
\usepackage{epsfig}

\usepackage{amssymb}

\usepackage{amsmath}
\usepackage{xspace}
\usepackage{rotate}
\usepackage[dvips]{color}
\usepackage{color}
\usepackage{latexsym}
\usepackage{txfonts}

\newcommand{\un}[1]{~\hspace{-1pt}\ensuremath{\mathrm{#1}}}
\newcommand{\titane}[1]{$\overset{^{{44}}}{}${#1}}

\newcommand{\integ}{{\it INTEGRAL}\xspace}
\newcommand{\ibis}{IBIS\xspace}
\newcommand{\spi}{SPI\xspace}
\newcommand{\isgri}{ISGRI\xspace}
\newcommand{\rosat}{{\it ROSAT}\xspace}

\newcommand{\asca}{{\it ASCA}\xspace}
\newcommand{\gro}{{\it GRO}\xspace}
\newcommand{\sax}{{\it BeppoSAX}\xspace}

\newcommand{\gammaray}{$\gamma$-ray\xspace}

\newcommand{\xray}{X-ray\xspace}

\def\cm3{cm$^{-3}$}

\def\etal{et~al.}
\def\ie{{\em i.e.~}}

\begin{document}

\begin{frontmatter}



\title{An \integ/\ibis view of Young Galactic SNRs through the $^{44}$Ti gamma-ray lines}


\author[1,2]{M. Renaud\corauthref{cor}},
\corauth[cor]{Corresponding author.}
\ead{mrenaud@cea.fr}
\author[3]{J. Vink},
\author[1,4]{A. Decourchelle},
\author[1,2]{F. Lebrun},
\author[2,1]{R. Terrier},
\author[1,4]{J. Ballet}

\address[1]{Service d'Astrophysique, DAPNIA/DSM/CEA, 91191 Gif-sur-Yvette, France}
\address[2]{APC-UMR 7164, 11 place M. Berthelot, 75231 Paris, France}
\address[3]{SRON National Institute for Space Research, Sorbonnelaan 2, 3584 CA Utrecht, The Netherlands}
\address[4]{AIM-UMR, 91191 Gif-sur-Yvette, France}


\begin{abstract}

We present preliminary results of \integ/\ibis observations on Cas~A, Tycho and Vela Junior Supernova remnants in the line emission of 
\titane{Ti}. This radioactive nucleus is thought to be exclusively produced in supernovae during the first stages of the explosion. It 
has a lifetime of about 87 y and is then the best indicator of young SNRs, as exemplified by the detection of \titane{Ti} 
in the youngest known Galactic supernova remnant Cas~A with \gro/COMPTEL and latter with \sax. In this paper, we will focus on this
SNR for which we confirm the detection of \titane{Ti} and point out the importance to know the nature of the hard \xray continuum, the 
Tycho SNR, for which no indication of \titane{Ti} was ever reported, and Vela Junior, for which the claimed detection of \titane{Ti} 
with COMPTEL is still controversial. The \integ/\ibis observations bring new constraints on the nature of these SNRs and on the 
nucleosynthesis which took place during the explosions.

\end{abstract}


\begin{keyword}

Gamma rays: astronomical observations \sep Gamma-ray sources (Cas~A, Tycho, Vela Junior) \sep 
Nucleosynthesis in supernovae \sep Supernova remnants in Milky Way
\PACS 95.85.Pw \sep 07.85.-m \sep 26.30.+k \sep 98.38.Mz

\end{keyword}

\end{frontmatter}




\section{Introduction}
\label{s:intro}

Supernovae (hereafter SNe) are the main galactic nucleosynthesis sites of production of radioisotopes which may be observed through 
their $\gamma$-ray line emission. Some of them are short-lived such as \titane{Ti}. The radioactive decay chain 
\titane{Ti}$\longrightarrow$\titane{Sc}$\longrightarrow$\titane{Ca}, with a half-life of about 60 yrs \citep{c:wietfeldt99}, 
produces three lines at 67.9\un{keV}, 78.4\un{keV} (from \titane{Sc}$^{\star}$) and 1157\un{keV} (from \titane{Ca}$^{\star}$) with 
similar branching ratios. This radioactive nucleus is thought to be created in all types of SNe but with a large variation of yields 
per type: from a few 10$^{-5}$ to $\sim$ 2 $\times$ 10$^{-4}$ M$_{\odot}$ for the most frequent SNe of Type II (Woosley \& Weaver, 
1995; Thielemann et al., 1996) and Type I$_{b/c}$ \citep{c:woosley95new} and up to 3.9 $\times$ 10$^{-3}$ 
M$_{\odot}$ for the rare event of the He-detonation of a sub-Chandrasekhar white dwarf (Woosley, Taam \& Weaver, 1986; Woosley \& 
Weaver, 1994). As reported by Iwamoto et~al.~(1999), the  \titane{Ti} yields for standard Type Ia SNe are between 8 $\times$ 10$^{-6}$ 
M$_{\odot}$ and 5 $\times$ 10$^{-5}$ M$_{\odot}$. It is primarily generated in the $\alpha$-rich freeze-out from nuclear statistical 
equilibrium occurring in the explosive silicon burning stage of core-collapse SNe, while a normal freeze-out Si burning is at play in 
Type Ia SNe (Thielemann, Nomoto \& Yokoi, 1986). Therefore, it probes deep into the interior of these exploded stars and provides a 
direct way to study the SN-explosion mechanism itself. On the other hand, it is strongly dependent on the explosion details, mainly on 
the mass-cut in core-collapse SNe (the mass above which matter is ejected), the energy of the explosion and asymmetries.

The \integ observatory \citep{c:winkler03} carries two main instruments: \ibis \citep{c:ubertini03} and \spi \citep{c:vedrenne03}. 
Both can provide images and spectra, based on the coded mask aperture system, working from 15\un{keV} to 1\un{MeV} and from 20\un{keV} 
to 8\un{MeV}, respectively. The line-sensitivity of the \ibis low-energy camera \isgri \citep{c:lebrun03} is really appropriate to 
detect the two low energy \titane{Ti} $\gamma$-ray lines at 67.9 and 78.4\un{keV} ($\Delta$E $\sim$ 6\un{keV} FWHM at 70\un{keV}). 
With a spectral resolution of $\sim$ 2\un{keV} at 1\un{MeV}, \spi can measure the ejecta velocity due to the Doppler broadening. 
We present here preliminary results on three young SNRs: Cas~A, Tycho and RX~J0852-4622 (Vela Jr).


\section{The Cassiopeia region: Cas~A and Tycho SNRs}
\label{s:Cas_reg}

\begin{figure}[htb]
\begin{center}
\includegraphics[scale=0.6, angle=0]{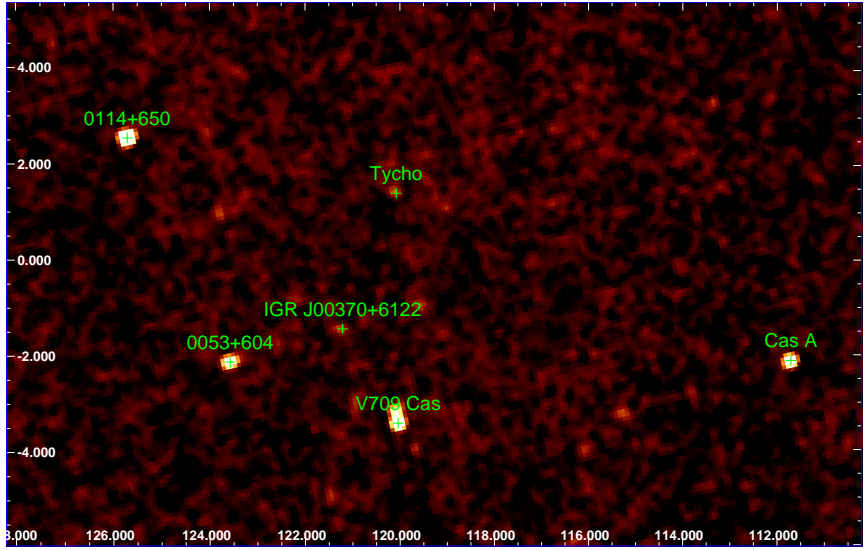}
\end{center}
\caption{A \ibis/\isgri image of the Cassiopeia region in the 25-40\un{keV} energy band.
         Cas~A and Tycho are detected at $\sim$ 25$\sigma$ and 6$\sigma$, respectively.\label{f:CasA_ima}}
\end{figure}

The Cassiopeia region was observed by \integ for a duration of $\sim$ 1.5 Ms. Figure \ref{f:CasA_ima} shows the region as observed by 
\ibis/\isgri in the 25-40\un{keV} band. Several sources have been revealed, amongst them Cas~A detected at $\sim$ 25$\sigma$ 
confidence level and Tycho SNR detected at $\sim$ 6$\sigma$ confidence level.

The discovery of the 1157\un{keV} \titane{Ti} $\gamma$-ray line emission from the youngest Galactic SNR Cas~A with COMPTEL
\citep{c:iyudin94} was the first direct proof that this isotope is indeed produced in SNe. This has been strengthened by the \sax/PDS 
detection of the two low-energy \titane{Ti} lines \citep{c:vink01}. By combining both observations, Vink et~al.~(2001) have deduced a 
\titane{Ti} yield of (1.5$\pm$1.0) $\times$ 10$^{-4}$ M$_{\odot}$. This huge value compared to those predicted by most of the models
could be due to several effects: a large energy of the explosion ( $\sim$ 2 $\times$ 10$^{51}$ erg), asymmetries \citep{c:nagataki98}
currently observed in the ejecta expansion, and a strong mass loss of the progenitor consistent with the scenario of a Type Ib SN
\citep{c:vink04}. In the case of Cas~A, the knowledge of the continuum emission is critical to properly measure the \titane{Ti} line 
flux. Unfortunately, it is still debated whether the nonthermal hard \xray continuum is synchrotron radiation or nonthermal 
bremsstralhung from supra-thermal electrons (see Vink 2005 for a recent review and references therein). 

\begin{figure}[htb]
\begin{center}
\includegraphics[scale=0.4, angle=0]{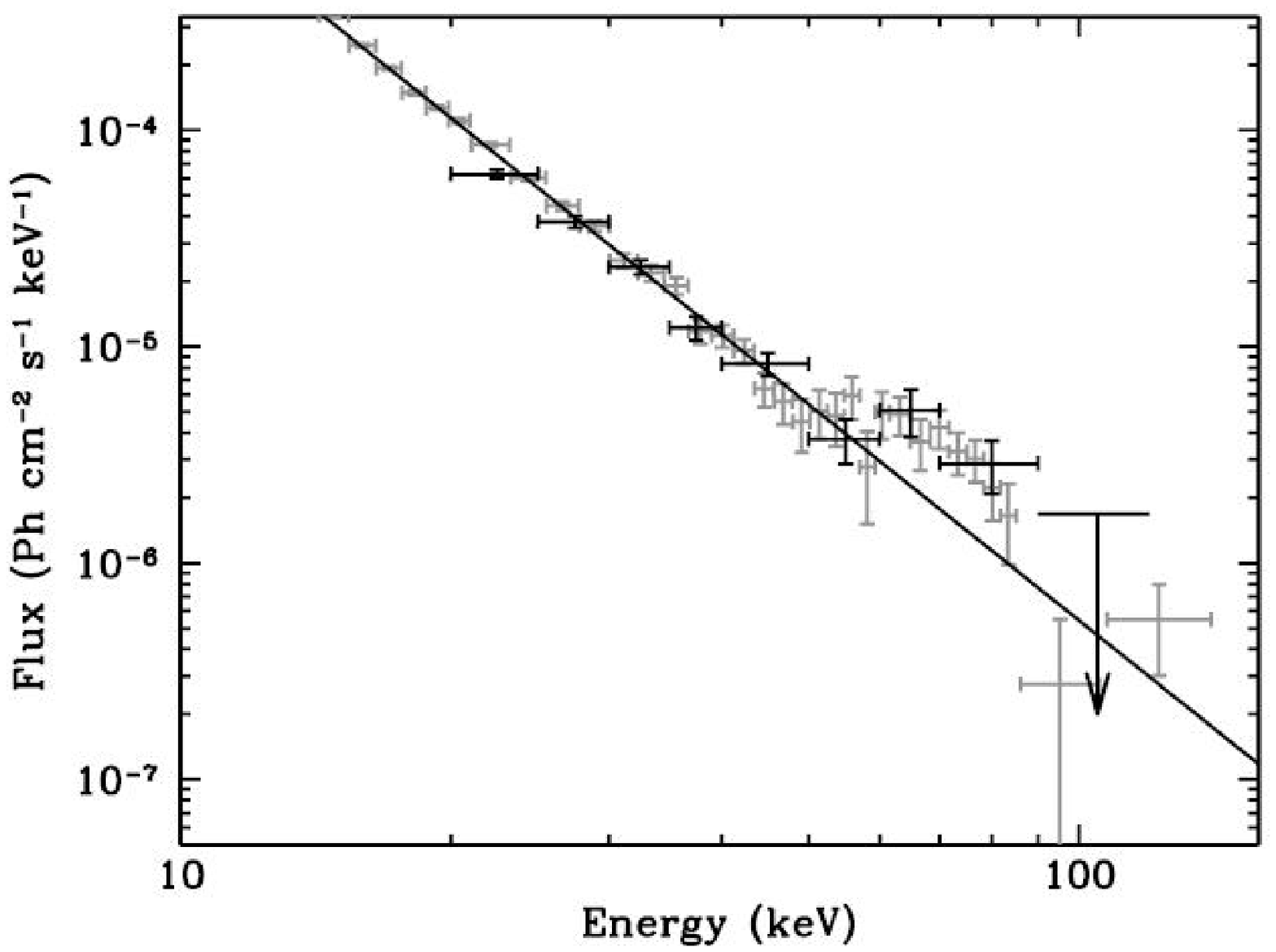}
\includegraphics[scale=0.3, angle=0]{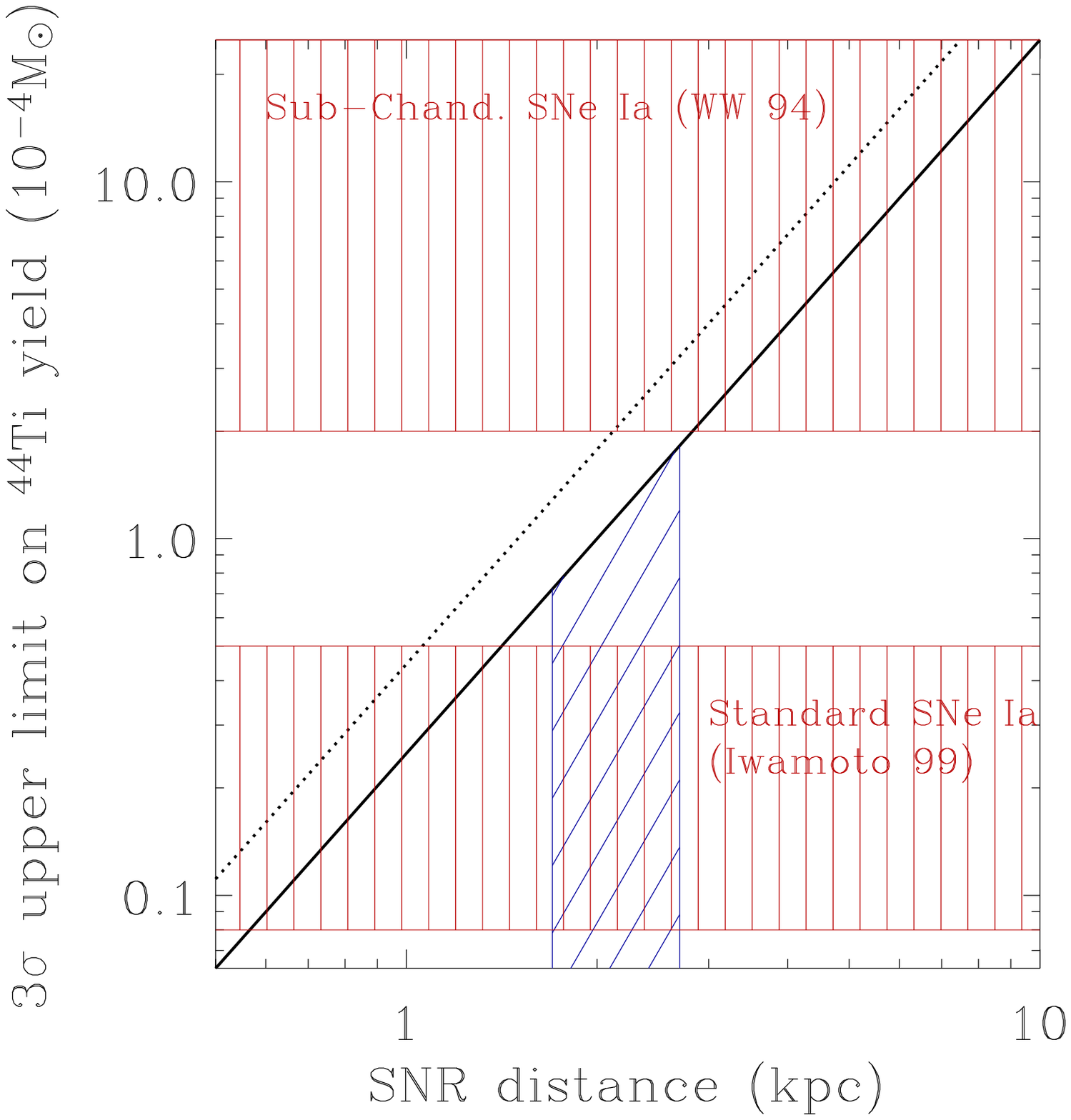}
\end{center}
\caption{(left) Spectrum of Cas~A obtained with \ibis/\isgri (in black) compared to
         that of \sax/PDS instrument (in grey). The continuum is assumed to be 
         a single power-law with a spectral index of $\sim$ 3.3. (right) 3$\sigma$ 
	 upper limit on the \titane{Ti} yield in Tycho as a function of the distance 
	 (black solid line). The dotted line corresponds to the \gro/COMPTEL results 
	 after the first 3 years \citep{c:dupraz97}. The two red areas represent 
	 the calculated yields for ``standard'' and Sub-Chandrasekhar He-detonation Type 
	 Ia SNe.\label{f:CasA_spe}}
\end{figure}

Figure \ref{f:CasA_spe} presents the spectrum obtained with \ibis/\isgri (in black, Vink 2005) compared to that of BeppoSAX/PDS (in 
grey). There is a 3$\sigma$ excess at the position of the first \titane{Ti} line with respect to a power-law continuum emission $\Gamma$ 
$\sim$ 3.3 (solid line). Both spectra are compatible, however, since there is still no clear detection of the continuum beyond the two 
\titane{Ti} lines, the weak S/N of the second line could be due to a steepening above $\sim$ 60\un{keV}, predicted for all synchrotron 
and some bremsstrahlung models. Assuming a power-law spectrum, the flux of the first \titane{Ti} line and that of each line by fitting 
both jointly are (2.3 $\pm$ 0.8) $\times$ 10$^{-5}$ cm$^{-2}$ s$^{-1}$ and (1.2 $\pm$ 0.6) $\times$ 10$^{-5}$ cm$^{-2}$ s$^{-1}$, 
respectively. By analyzing the \spi data, we didn't find any excess neither in the broad (1142 - 1172\un{keV}) nor in the narrow energy 
band around the 1.157\un{MeV} \titane{Ti} line yielding to a preliminary 2$\sigma$ lower limit on the ejecta velocity $\Delta$v 
$>$ 10$^{3}$ km s$^{-1}$ for an assumed line flux of 1.9 $\times$ 10$^{-5}$ cm$^{-2}$ s$^{-1}$.

The Tycho SNR is the prototype of a Type Ia SN \citep{c:baade45}. No evidence of \titane{Ti} has ever been reported 
\citep{c:dupraz97}. With an age of 433 yr and a distance of 2.2 $\pm$ 0.5 kpc, this SNR is the most promising candidate to observe 
explosive nucleosynthesis products of thermonuclear SNe. As shown in Figure \ref{f:CasA_ima}, this SNR is detected by \isgri in the
hard \xray continuum up to $\sim$ 50\un{keV} but we didn't find any significant excess in the range of the two low energy \titane{Ti}
lines. Our 3$\sigma$ upper limit of 1.5 $\times$ 10$^{-5}$ cm$^{-2}$ s$^{-1}$ can be translated into an upper limit on the \titane{Ti}
yield. Figure \ref{f:CasA_spe} shows this value as a function of the distance of the SNR. One can see that all the models of 
Sub-Chandrasekhar Type Ia SNe, predicting huge \titane{Ti} yields, are excluded for any distance inside the uncertainties. On the other
hand, we cannot at this time really constrain the ``standard'' Type Ia models exposed by Iwamoto et~al.~(1999). Further results on 
these two SNRs based on a significantly longer observing time ($\sim$ 3 Ms) are expected in the near future.


\section{Vela Junior}
\label{s:VelaJr}

Since its detection with \rosat and COMPTEL in the Vela region, RX~J0852-4622 (Vela Jr) is still a mystery. Previous estimates based 
on its apparent diameter ($\sim$ 2$^{\circ}$), the spatially coincident excess in the 1.157\un{MeV} \titane{Ti} line, and the \rosat 
\xray spectrum have showed that this SNR is likely young ($\sim$ 700 yr) and nearby ($\sim$ 250 pc). However, the relative strong 
absorption observed by \asca towards the source and the weak radio flux support a ``not so nearby, and so, not so young'' scenario. 
Moreover, the re-analysis of the COMPTEL data found that the detection of this SNR as a \titane{Ti} source is only significant at the 
2-4$\sigma$ confidence level. Surprisingly, Tsunemi et~al.~(2001) and Iyudin et~al.~(2005) have detected a feature in the \xray 
spectrum at $\sim$ 4.1\un{keV} which could come from Ti and Sc excited by high velocity collisions in the SNR outer shell. Iyudin 
et~al.~(2005) argued that the consistency of this \xray line flux and the 1.15\un{MeV} \titane{Ti} line flux seems to support the 
first estimations of age and distance. \integ has deeply observed this region during the two first years. We have analyzed data in the 
range of the two low energy \titane{Ti} lines but we did not find any evidence of \titane{Ti}. Our non-detection could be compatible 
with the COMPTEL findings if Vela Jr appears as an extended source for the \ibis telescope: in that case, the total flux should be 
diluted over all the sky pixels and then in any direction within the remnant, the flux would go below our sensitivity. Our 3$\sigma$ 
upper limit is close to one fourth of the COMPTEL flux and then we can exclude four separated point-like sources ($\Phi$ $<$ 8') with 
the same flux inside the remnant (\ie a scenario where the \titane{Ti} would be located in ``hot-spots''). A method to reconstruct the 
flux of an extended source with a coded mask telescope is under study \citep{c:renaud06}.


\vspace{0.5cm}
\begin{table}[h]
\begin{center}
\begin{tabular}{|c|c|c|}
\hline
SNR		&	ISGRI									&	SPI \\
\hline
Cas A		&	2.3$^{\pm 0.8}$ $\times$ 10$^{-5}$ cm$^{-2}$ s$^{-1}$ (67.9\un{keV})	&	$<$ 3.1 $\times$ 10$^{-5}$ cm$^{-2}$ s$^{-1}$ \\
\hline
Tycho   	&	$<$ 1.5 $\times$ 10$^{-5}$ cm$^{-2}$ s$^{-1}$ (67.9 \& 78.4\un{keV})	&	? \\
\hline
Vela Jr. 	&	$<$ 10$^{-5}$ cm$^{-2}$ s$^{-1}$ (67.9 \& 78.4\un{keV})			& 	$<$ 1.1 $\times$ 10$^{-4}$ cm$^{-2}$ s$^{-1}$ (78.4\un{keV}) \\
\hline

\end{tabular}
\vspace{0.3cm}
\caption{Summary of the preliminary results obtained with \integ in the range of the 
\titane{Ti} \gammaray lines. The upper limits were calculated assuming that sources appear
\underline{as point-like} and are given at the 3$\sigma$ confidence level.\label{tab_sum}}
\end{center}
\end{table}


\section{Discussion}
\label{s:discuss}

This paper summarized our preliminary results on young SNRs through the \titane{Ti} \gammaray lines. One of the main goal within this 
framework is the search for ``young, missing, probably hidden'' SNRs. The non-detection by HEAO-3, SMM and recently COMPTEL of any 
such sources in the inner part of the Galaxy seems to be incompatible with what we expect to see from 3 SNe per century, most of them 
core-collapse SNe, producing $\sim$ 10$^{-4}$ M$_{\odot}$ of \titane{Ti}, as observed in Cas~A or derived for SN~1987A. On the other 
hand, current nucleosynthesis models can only explain one third of the solar abundance of \titane{Ca} \citep{c:timmes96}, thought to 
come mainly from the radioactive decay chain of the \titane{Ti}. We also performed a first analysis of the Galactic Central regions 
with \ibis/\isgri and confirm that there is no evidence of any strong excess, \ie a young SNR \citep{c:renaud04}. In any case, these 
first results show that we can study \gammaray lines with the \ibis/\isgri with a line sensitivity after only two years of observation 
better than those of previous \gammaray instruments and then bring new constraints on the explosive nucleosynthesis production in SNe. 

M.R. gratefully thanks the organizers for this nice meeting. The present work was based on observations with \integ, an ESA project with 
instruments and science data center (ISDC) funded by ESA members states (especially the PI countries: Denmark, France, Germany, Italy, 
Switzerland, Spain, Czech Republic and Poland, and with the participation of Russia and the USA). \isgri has been realized and 
maintained in flight by CEA-Saclay/DAPNIA with the support of CNES.



\end{document}